\documentclass[letterpaper, 10 pt, conference]{ieeeconf}  
\let\labelindent\relax
\usepackage{todonotes}

\IEEEoverridecommandlockouts                              

\usepackage[puttinydots,useemptybox]{braille}
\usepackage{wrapfig}
\usepackage{enumitem}
\usepackage{times}
\usepackage{graphicx}
\usepackage{listings}
\usepackage{amsmath}
\usepackage{amssymb}
\usepackage{caption}
\usepackage{subcaption}
\usepackage{babel, blindtext, amsmath}
\usepackage[margin=0.8in]{geometry}
\usepackage{csquotes}
\usepackage{cite}
\usepackage{listings}
\usepackage{xcolor}
\usepackage{hyperref}
\hypersetup{
    colorlinks=true,
	citecolor=blue,
    linkcolor=blue,
    filecolor=magenta,      
    urlcolor=blue,
}

\definecolor{codegreen}{rgb}{0,0.6,0}
\definecolor{codegray}{rgb}{0.5,0.5,0.5}
\definecolor{codepurple}{rgb}{0.58,0,0.82}
\definecolor{backcolour}{rgb}{0.95,0.95,0.92}

\lstdefinestyle{mystyle}{
    backgroundcolor=\color{backcolour},   
    commentstyle=\color{codegreen},
    keywordstyle=\color{magenta},
    numberstyle=\tiny\color{codegray},
    stringstyle=\color{codepurple},
    basicstyle=\ttfamily\footnotesize,
    breakatwhitespace=false,         
    breaklines=true,                 
    captionpos=b,                    
    keepspaces=true,                 
    numbers=left,                    
    numbersep=5pt,                  
    showspaces=false,                
    showstringspaces=false,
    showtabs=false,                  
    tabsize=2
}

\title{Vibration-based communication for deafblind people}

\author{David C. Kutner$^1$ and Suncica Hadzidedic$^1$
\thanks{$^1$Department of Computer Science, Durham University, UK {\tt\small david.c.kutner@durham.ac.uk}}%
}

\begin{document}
\emergencystretch 3em 

\maketitle

\begin{abstract}
Deafblind people have both hearing and visual impairments, which makes  communication with other people often dependent on expensive technologies e.g., Braille displays, or on caregivers acting as interpreters. 
This paper presents Morse I/O (MIO), a vibrotactile interface for Android, evaluated through experiments and interviews with deafblind participants. MIO was shown to enable consistent text entry and recognition after only a few hours of practice. The participants were willing to continue using the interface, although there were perceived difficulties in learning to use it. Overall, MIO is a cost-effective, portable interface for deafblind people without access to Braille displays or similar. 
\end{abstract}

\section{Introduction}

Deafblind people have both vision and hearing impairments resulting in significant communication challenges in their day-to-day lives. The World Federation of the Deafblind (WFDB) has estimated that 0.21\% of all individuals aged five and over have severe deafblindness, while noting that there is no agreement on ``best practice" definitions or measurements of deafblindness \cite{wfdb2018global}. 

 Medical journals \cite{moller2003deafblindness} emphasize the importance of combating social isolation among deafblind people. A survey of 28 deafblind people from Europe \cite{hersh2013deafblind} found that many had a strong desire to contribute to society, and were active in different organizations, including in leadership roles. However, assistive devices for deafblind communication - e.g. Braille, hearing aids,  text-to-speech - require further improvement \cite{hersh2013deafblind}. High costs are a key factor restricting access to assistive devices \cite{hersh2013deafblind}. The WFDB emphasizes that a lack of access to affordable assistive technology and support contributes to deafblind people' ``poverty trap" \cite{wfdb2018global}.  

\begin{figure}[!h]
\centering

\includegraphics[width=0.4\textwidth]{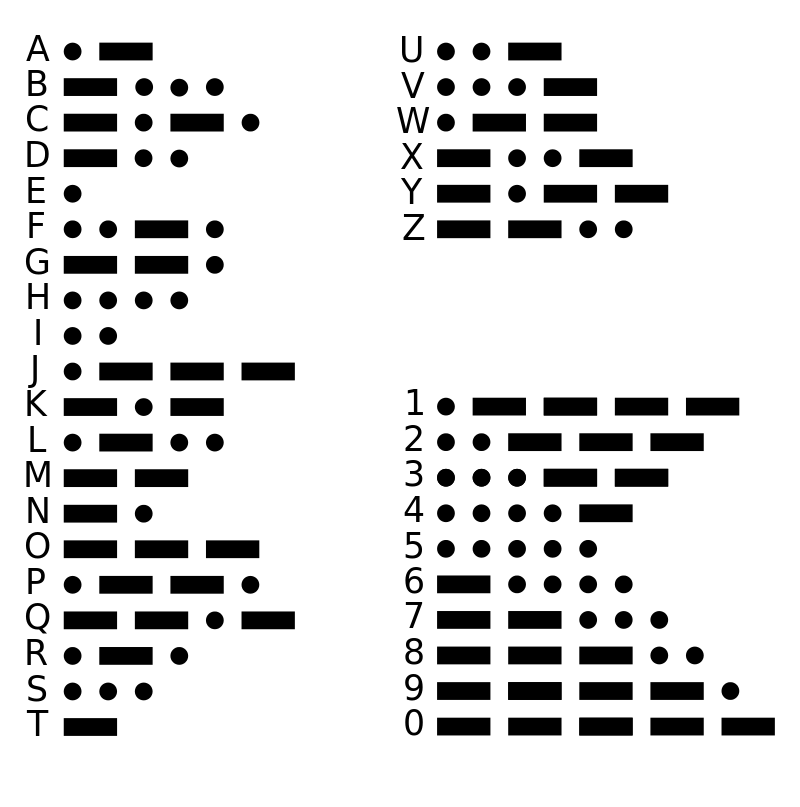}
\caption{Chart showing the Morse encodings of the alphabet and decimal numbers per the International Telecommunication Union (ITU) standard \cite{itu2012morse}}
\label{fig:internationalmorsecode}
\vspace{-12pt}
\end{figure}

This paper aims to present a low-cost, easy-to-learn vibrotactile interface for deafblind users' interaction with computers. It uses Morse code - the encoding of the Latin alphabet and special characters as sequences of dots and dashes (examples in Fig. \ref{fig:internationalmorsecode}) \cite{itu2012morse}. The Morse Input/Output (MIO) interface enables entry of strings of any length into any text field on an Android device, with haptic feedback to users. A client-side application was developed as a learning environment and for gathering data about system usage.

The principle of ``Nothing about us without us" \cite{charlton2000nothing} promotes the inclusion of people with disabilities in decision-making processes in any area that has an impact on their lives. In keeping with this, deafblind users evaluated the MIO application. The design, implementation, and evaluation processes in this study were informed by feedback from over two dozen organizations, academics, and professionals. 

Overall, the main contributions of this paper are: the MIO vibrotactile interface, interface evaluations specifically with deafblind users, and an evaluation of practical implications about MIO's usability and real-world applicability.

\subsection{Related Work} \label{section:related work}

Applications of haptic devices designed for deafblind people have been the focus of some research in the last decade. However, due to the difficulties with recruiting deafblind people, these studies often had sighted and/or hearing participants evaluate the technology in question \cite{caporusso2017wearable} \cite{plaisier2020learning} \cite{norberg2014web} \cite{ranjbar2009vibrotactile}. For example, a 2020 study \cite{plaisier2020learning} aimed to evaluate the learnability of Morse code without a visual reference. The study's 10 participants (all with normal vision)  succeeded in learning at least 15 (and up to 24) letters in Morse code within just 30 minutes. Noteworthy is the study's method of introducing a small group of letters at a time and teaching the recognition of words composed of these letters before the entire alphabet is learned. However, this study was limited to feedback only from sighted participants, with dots and dashes applied to the left and right arms, respectively, a technique that is not possible to implement on a single phone. 

Implementation of a vibrotactile Morse-based system for deafblind communication has already been researched. In 2014, Norberg et al. \cite{norberg2014web} developed a web navigation system using an Xbox 360 controller to ``read" text to a user operating an otherwise unmodified computer. The study was carried out with ``simulated" deafblind participants, who wore noise-canceling headphones and blindfolds. They found that three out of four participants were successful in completing a series of simple navigation tasks with just 15 minutes of training in Morse code. The study's conclusions were undermined by the subjects' likely prior experience using a mouse and cursor. 

Samsung Good Vibes is an application for deafblind communication that has been available since 2019 \cite{akkara2019smartphone}.  It has two interfaces - one for the deafblind, the other for sighted users. It allows for two-way communication, much as a messaging application would, using vibrotactile Morse to relay message contents to a deafblind user. There are no sources confirming its use outside Asia, and unfortunately, at the time of writing, no studies on usability or learnability of this App were publicly available.

In 2014, Arato et al. \cite{arato2014teaching} developed, for a single deafblind individual (PG), an interface for communicating by SMS on an Android phone. Because PG was already familiar with Braille, he was enabled to enter text in Braille. However, recognizing Morse characters by vibrations was more difficult. PG was able to sense 15 characters per minute, or 10 characters per minute for unknown texts, and continued to use his Morse-Braille vibrating phone in his day-to-day life after the study. This hybrid Morse-Braille approach was considered to be low-cost and usable.

Overall, there has been scant research into vibrotactile assistive technology for the deafblind. The only reference to an interface implemented on a smartphone, using Morse code for text entry, was Samsung Good Vibes \cite{akkara2019smartphone}, although no experimental results were available. Common to all of these studies, with the exception of \cite{arato2014teaching}, is the absence of deafblind people from the design and evaluation process.

\section{Morse Input/Output} \label{solution}

In this section, we present the feedback from professionals and academics and the design and development of the MIO interface. 

\subsection{Interviews with professionals} \label{subsection:interviews}

\begin{displayquote}
\textit{``The average person pays \$400 to \$1000 for a mobile phone - I don't mind paying that, but my Braille display is a minimum \$1800 to as much as \$5000."} - Scott Davert, 2021
\end{displayquote}

Our study applied the user-centered design approach \cite{abras2004user}. Academics and professionals who have worked with deafblind people thus participated in the design and evaluation of the MIO application.

\textbf{Dr. Marion Hersh} (Glasgow University) has published over a dozen papers on deafblindness and assistive technology, for which she interviewed deafblind people across Europe \cite{hersh2013deafblind}. In a discussion for this paper, Dr. Hersh raised the concern that ``Morse has been developed for convenience of transmission rather than convenience of understanding", as well as the possibility that longer patterns (i.e., more dots and dashes) might be harder to learn. Consequently, in the experiment, we  used more complex letters (e.g., C, -•-•) as well as concisely encoded ones (e.g., A, •-). She also acknowledged the benefits of a smartphone application because of the wide availability of smartphones, which are more affordable and also more discreet than assistive devices for use in public.

\textbf{Heather Colson-Osborne} has 13 years of experience at the Anne Sullivan Centre in Ireland, which supports deafblind individuals. She interpreted for one of our participants and contributed to the design of the interface. The idea of ``touch recording" was introduced as the Centre's staff pointed out that many deafblind people are illiterate, having learned a limited tactile vocabulary, e.g. a spoon to refer to mealtime. With touch recording, a user could tap or hold their finger to the screen, and the smartphone would register the timings and replay the message as vibrations of the lengths of time pressed.

\textbf{Scott Davert} has previously worked with hundreds of blind and deafblind people as a Braille teacher and the coordinator for New York State's federally funded iCanConnect program. He is himself deafblind and proficient in Morse code, thus uniquely suited as a contributor to this project. He was particularly interested in integrating MIO as an alternative interface to the main method on his phone. On Android devices, the only way to connect a Braille display is through Bluetooth, which is not consistently reliable. According to Mr. Davert, a Morse vibration-based interface would give deafblind people ``a chance to repair [their] device without having to ask someone who can either see or hear to do it. It's something I would definitely use".

Mr. Davert further explained Android's lack of popularity among blind people: ``The Braille
access is bad - it’s really bad. And they know; I actually went to Mountain View [...] and told them!". The accessibility features of the latest Android version are comparable to those of iOS 4. Therefore, he warned that it would be very challenging to find deafblind participants with access to an Android phone. Nevertheless, for the purposes of MIO, he acknowledged that it might be more fruitful to integrate the interface into an open-source Android operating system than iOS, which has major restrictions for new software development.

\subsection{MIO design}
Based on the interviews with professionals detailed above, MIO was designed for individuals whose disabilities range from total deafness and blindness to a severe impairment in hearing and vision. A Graphical User Interface (GUI) was included because some deafblind users have partial vision, and to facilitate development. MIO is an interface which uses vibrations for text output and a large keyboard for text input. It was implemented in Java for Android OS.

Although the professional participants' input (Section II-A) implied that deafblind people are more likely to own iOS devices, an Android platform was chosen for MIO, as it is more conducive to receiving experimental software. Moreover, bespoke setups, e.g. using a Raspberry Pi, were excluded for reasons of cost. Cost is an important factor restricting access to assistive technology among deafblind people \cite{wfdb2018global} \cite{hersh2013deafblind}. For this study, which was based on user-centered design \cite{abras2004user}, one of the crucial requirements was accessibilty of the application.  

Finally, we have chosen Morse code over an encoding derived from Braille (or a hybrid one, as in \cite{arato2014teaching}) primarily for the conciseness and simplicity of the Morse alphabet. Although Braille is much more popular among blind people, it is more complex and less time-efficient for use through a touchscreen. The number of dots in a Braille character depend on its position in the alphabet. The letter T, for instance, is a single dash in Morse; it is represented by \braille{t} in Braille, and would therefore require more time to both read and input.

\subsubsection{The MIO interface} 

\begin{figure}[!h]

\centering
	\begin{subfigure}[b]{0.48\linewidth}
		\centering
		\includegraphics[width=\textwidth]{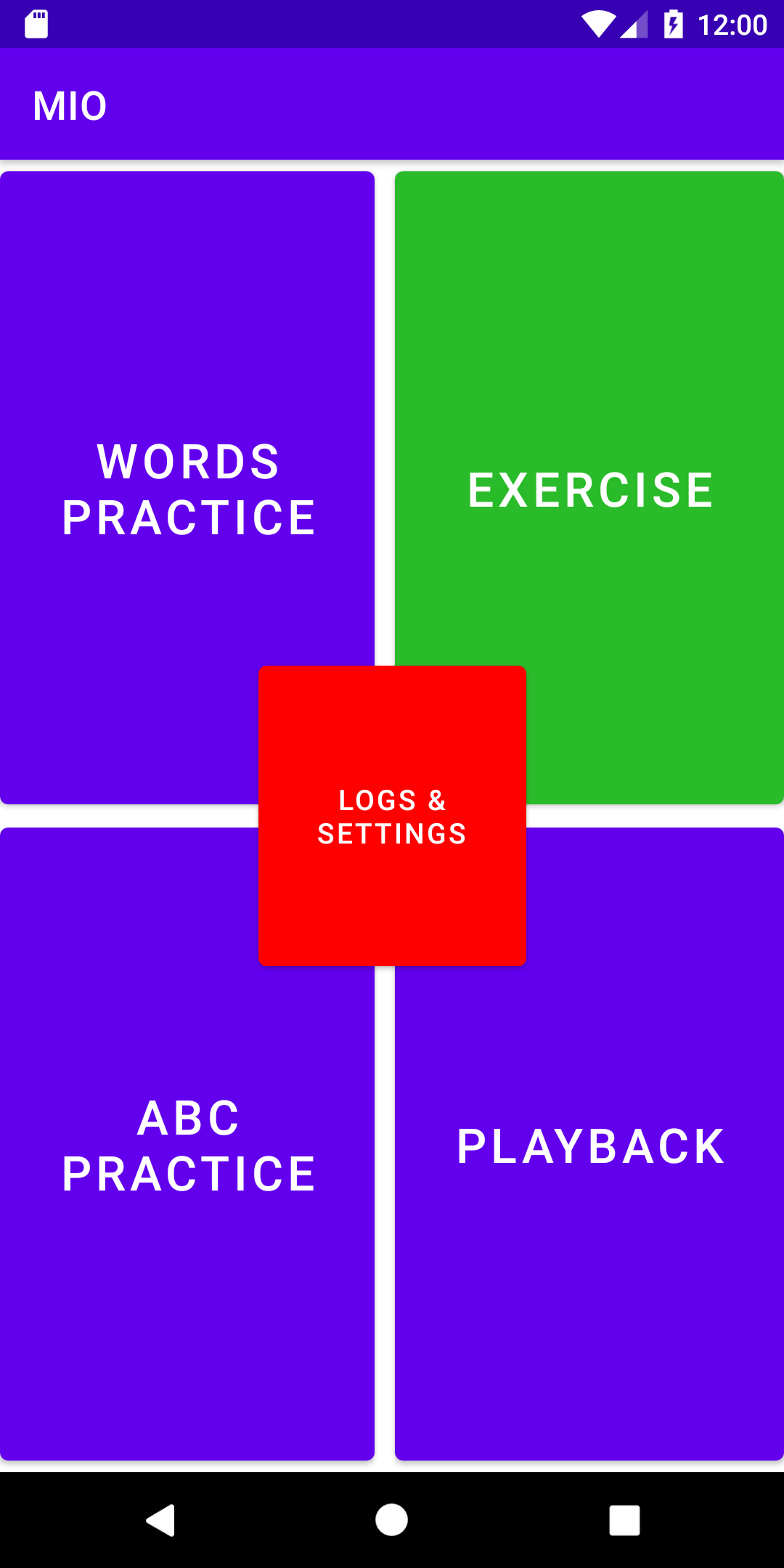}
		\caption{Landing screen.}
		\label{fig:landing}
	\end{subfigure}
	\hfill
	\begin{subfigure}[b]{0.48\linewidth}
		\centering
		\includegraphics[width=\textwidth]{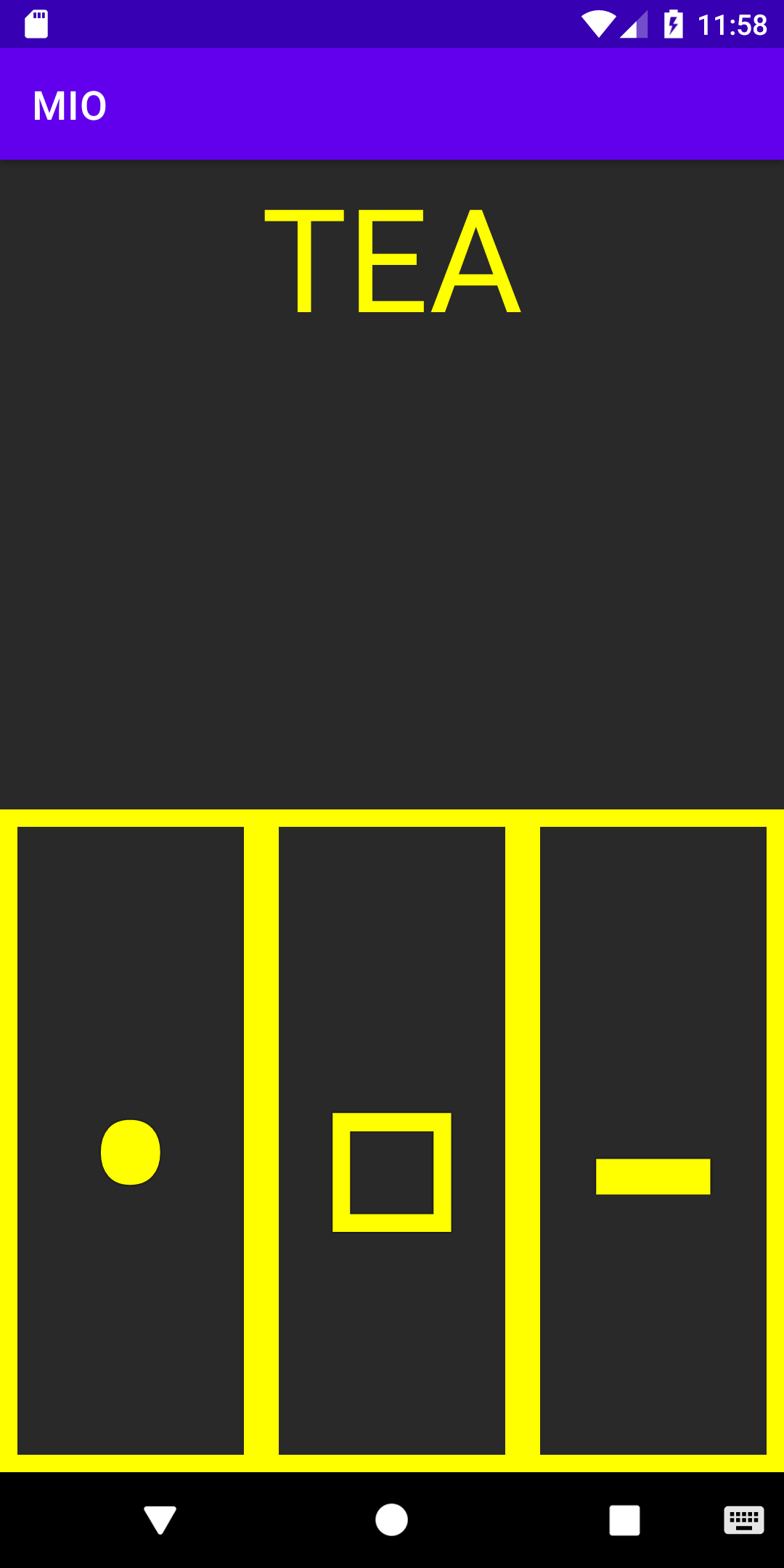}
		\caption{Words practice.}
		\label{fig:tea}
	\end{subfigure}
	\caption{The MIO application on Android.}
	\label{fig:app}
\end{figure}

The MIO interface consists of an \textbf{output method} - Morse vibrations, and an \textbf{input method} - the MIO keypad. A brightly-colored GUI (Fig. \ref{fig:app}) is also implemented for ease of use for the partially sighted.

\textbf{Morse vibrations}. Text output is performed by emitting vibrations of different lengths, directly applying Morse code according to the ITU standard \cite{itu2012morse} (Fig. \ref{fig:internationalmorsecode}). After some calibration with a user, the length of one unit (a dot) was set at 200 ms.  Thus, e.g., the word “TEA”, which is encoded as -\hspace{10pt}•\hspace{10pt}•- in Morse, is emitted as: 600 ms vibration (-); 600 ms pause (between letters); 200 ms vibration (•); 600 ms pause (between letters); 200 ms vibration (•); 200 ms pause (within a letter); 600 ms vibration (-).

\textbf{MIO keypad.} The input method is a keypad divided horizontally into three buttons on the bottom half of a smartphone screen (Fig. \ref{fig:tea}). From left to right, the buttons are dot, square and dash. The button outlines and the symbols are in yellow to maximize readability \cite{legge1990psychophysics} for use by partially sighted persons and  sighted carers. 
We added a square key for ease of use and to minimise input mistakes. The square key is pressed once to ``submit" a letter, twice to enter a space, and three times to create a new line. A user wishing to type the phrase ``HI PAL" would thus enter the sequence \texttt{••••$\square$••$\square\square$•--•$\square$•-$\square$•-••}.
For dots and dashes, the length of input and output vibrations is identical. Pressing the dot key produces a 200 ms vibration, while the dash key vibrates for 600 ms. When pressed, the square button vibrates for 100 ms, the shortest length  perceptible by our participants. Thus, all keys produce unique haptic feedback, so that users can  easily differentiate between the buttons. Swiping the keyboard from right to left resets the current input, emitting a 1200 ms vibration to distinguish the action from an accidental button touch.

\subsubsection{MIO activities}

The MIO application includes several activities intended to familiarize users with the interface. Fig. \ref{fig:app} shows the landing screen (Fig. \ref{fig:landing}) and a \textbf{practice activity} (Fig. \ref{fig:tea}). The layout is intended to enable a deafblind user to easily find buttons from the edge of the screen. 

We included a unique vibration that is emitted at the start of an activity, at the suggestion of our participants. One 100 ms vibration is emitted for Words Practice, two for Exercise, three for ABC Practice, and four for Playback. Swiping backwards from right to left in any activity  returns the user to the main screen. Navigating to the main screen emits the string ``MIO" in Morse. 

The \textbf{Settings activity} is intended for a sighted helper. It enables access to the system logs and customization. The \textbf{Words Practice}, \textbf{ABC Practice} and \textbf{Exercise} activities are all 
\textbf{Practice activities} (Fig. \ref{fig:tea}), for learning to recognize and type a prompt in Morse. Tapping the top half of the screen vibrates a word displayed there. Successfully entering the word progresses the user to the next word. An incorrect entry resets the input and emits a 1200 ms vibration to prompt the user to re-attempt entry. 

To learn and practice individual letters, users can open the \textbf{ABC Practice} activity, which cycles through the alphabet. The \textbf{Exercise} activity presents a specially designed series of prompts building up to the phrase ``THE CAT EATS THE CAKE". This design, using just 6 letters, is in line with the Plaisier et al's \cite{plaisier2020learning} approach of first teaching users a small number of letters so that they can read and enter entire words, before they learn and memorize all the letters of the alphabet. 

The \textbf{Words Practice} activity, by default, displays random words from Basic English \cite{ogden1935basic}. However, it can be configured in the settings. Consequently, users can practice through arbitrary lists of strings and learn Morse code by, e.g., using a poem they are already familiar with. 

Finally, with the \textbf{Playback} activity, users can go through the alphabet without typing it in Morse. Simply tapping the screen progresses the user to the next prompt and emits it as a Morse vibration. The application can also be configured to playback lines from a familiar text. 

\section{User Studies}

One of the main contributions of this paper was that we specifically recruited deafblind people to participate in the design and evaluation of the MIO application. User-centered design \cite{abras2004user} ``assures that the product will be suitable for its intended purpose in the environment in which it will be used". The selection of participants was reinforced by the feedback received in the pre-study interviews with professionals (\ref{subsection:interviews}). Moreover, the sampling decision was consistent with the literature on assessment of assistive device usability for people with disabilities \cite{lazar2017research}, and the ``Nothing about us without us" approach promoted by their community \cite{charlton2000nothing}. 

Therefore, a purposive sampling approach \cite{etikan2016comparison} was adopted. We reached out to more than 90 deafblind organizations worldwide, of which 26 responded. However, in the final study, only two deafblind individuals (AC and WK) participated, both recruited through the Anne Sullivan Foundation in Ireland.

Particular attention was given to protecting the participants' privacy and obtaining their informed consent for data collection. The project received ethics approval from the Durham University Department of Computer Science.
Both participants agreed to having their interviews video-recorded and to reporting of their direct quotes. Each video call lasted a few hours, with approximately one week between calls (i.e., study stages).  

Interview and experiment design was semi-structured. The small number of participants made it possible to change the order of questions and tasks on-the-fly, tailored to the interests and engagement of the participants. Semi-structured interviews with an open response format grant both the assurance that relevant topics are covered and the flexibility to allow participants to point out issues  \cite{hersh2013deafblind}.

The user study had \textbf{three stages}:
\begin{enumerate}
\item \textbf{Background and preliminary questions.} The participants were asked to report on: their experience with deafblindness (e.g., at what age their impairments developed); methods of communication they had previously used; assistive technologies they were familiar with, and how these might be improved. The informed consent to being recorded for the study was also gathered at this stage, and the correct configuration of the application was verified.

\item \textbf{Application description, experiment, and evaluation.} In the second stage, the MIO application was presented. The interviewer provided instructions directly to one participant (WK), as the MIO application was designed for use and navigation without visualization of the screen. The other participant (AC) received instructions through her interpreter.

First, the \textbf{ABC Practice} activity was used to demonstrate the letter A output as vibrations and how to input it. Participants were then encouraged to continue practicing with successive letters until comfortable to attempt the \textbf{Experiment} activity. The experiment required attempting a sequence of prompts, building up to the sentence ``THE CAT EATS"; starting with one letter at a time, then whole words, and finally the sentence itself, using 5 letters of the alphabet.

Following the experiment, the MIO application was evaluated. Participants rated (Likert scale: 1 - strongly disagree to 5 - strongly agree) the generic 10 questions from the System Usability Scale (SUS) \cite{brooke1996sus} and elaborated on their answers.

We also asked an additional set of questions, based on SUS, however, tailored to the MIO application. These were:
\begin{enumerate}[label=(\arabic*)]\itemsep=0pt
\item	I understand how the app is intended to work; it was easy to conceptualize it.
\item	Typing words using this method is physically difficult. 
\item	I could find keys without issue.
\item	I often pressed the wrong button by accident.
\item	It was always immediately clear to me what button I had just pressed.
\item	The idea of learning Morse code seems easy to me.
\item	I often had to remind myself of what I had to type by tapping the top of the screen.
\end{enumerate}
At the end of the experiment, the participants' ability to recognize and recall letters was tested. They were asked to play the letter A vibration, then to give the Morse representation of letter C.
Lastly, they evaluated the MIO interface by reporting on: the perceived speed of communication in comparison with other means of communication; potential obstacles to using the interface; ways to usefully integrate MIO to facilitate day-to-day activities; and perceived ease of learning Morse code. As a preparation for the follow-up interview, participants were encouraged to practice using the MIO interface. 

\item \textbf{Follow-up interviews.} A week later, participants were asked to reflect on the frequency of MIO use since the last interview, to test if their recognition of letters in vibrotactile Morse improved. The participant WK re-attempted entering the phrase ``THE CAT EATS", and completed the SUS questionnaire again.

\end{enumerate}

\section{Results}

This section presents qualitative and quantitative results from the MIO interface evaluation. 

\subsection{Participant background information}

\textbf{Participant AC}, a 50-55 year old female, was born deaf, and lost her vision at the age of 12. She learned to lipread and Irish Sign Language (ISL) at school as a child. She also knows and uses Braille. AC normally lives at a care center. She has a support worker for daily interactions and assistance in using her phone. Therefore, interviews with AC were relatively slow due to the delay in interpretation between English and ISL. Moreover, there were difficulties in conveying certain concepts, such as the level of agreement on a numerical scale. According to her support worker, answering questions in this manner was completely new to her.

\textbf{Participant WK}, a 20-25 year old female, had 10\% vision from birth; she was able to read text and see pictures with a magnifier, but could only see nearby objects. At the age of 10, her vision deteriorated, hence she had to learn to read Braille. Her hearing was above 60\% until the age 17. It is now at 0\%, although she can still communicate through speech by using special hearing aids. WK is proficient with assistive technology; she makes use of both Braille displays and text-to-speech technologies in her day-to-day life, enabling her to navigate complex systems. However, she said she experienced frustration as the software that she needed often lacked accessibility features. 

\subsection{MIO usability results}

Participant WK provided two sets of answers. The first time she used the system, she evaluated it with an SUS score of 57.5. After four days of further practice, she was more familiar with the system, and rated it with an above average SUS score of 75 (SUS = 68 is an average experience, and SUS=80 is a good experience \cite{lewis2018item}). 
Participant AC's numerical scores were estimated with the help of her support worker on the basis of natural-language (ISL) responses to the prompts. Only one set of answers was collected, with a low total score of 10  \cite{lewis2018item}. 
Participants' detailed feedback to the questions is summarized below. 

\textbf{Participant AC}:
\textit{``If staff help me, I can learn to do it."} 

AC was at times frustrated and confused by the questionnaire, possibly more so than by the MIO application. AC would like to use MIO ``a little bit" going forward. She considered that it had a steep learning curve, and that most people would be slow to adapt to it. For her, the MIO interface was ``hard to use," and she would need staff to help her. Her support worker reported she had difficulties finding and pressing the right keys at first, although this improved over the course of the experiment. AC had difficulty counting vibrations and differentiating between dots (200 ms) and dashes (600 ms). Nevertheless, she was enthusiastic to participate in the study and eager to learn. Upon her request, the experiment was modified on-the-fly to allow her to feel a word she chose. In the end she expressed the willingness to continue learning to use the MIO interface. 

\textbf{Participant WK}
\begin{displayquote}
\textit{I feel very confident because I know where everything is.} 
\end{displayquote}
WK evaluated the MIO application twice. After the first experiment, WK reported: ``kind of 50-50, but once you start using it you get quicker". She added that she would be using the application more after the experiment, and emphasized that it is important to have ``a bit of time to get used to it". In the second evaluation, WK's answers implied an improvement to the application's perceived usability.

WK found the MIO application broadly understandable, consistent, well-integrated, and not cumbersome. However, she noted that the application's feedback was unclear. At WK's request, the application was modified to only play the suffix to the current input, enabling the user to quickly identify the next letter in the prompt.

Overall, WK expressed an enthusiasm to continue using the MIO interface. She found it would be useful as a fallback method for answering messages, for instance if a Braille display was not available. She perceived the MIO's haptic feedback as very useful for avoiding typos and checking input. 

\subsection{System log analysis}

The MIO log included the following data about each prompt attempt: time, whether it was successful, and playback frequency. AC attempted 23 prompts during a single experiment, while WK attempted 74, over two experiments and in her own time.

\begin{figure}[!h]
\centering
\includegraphics[width=\linewidth]{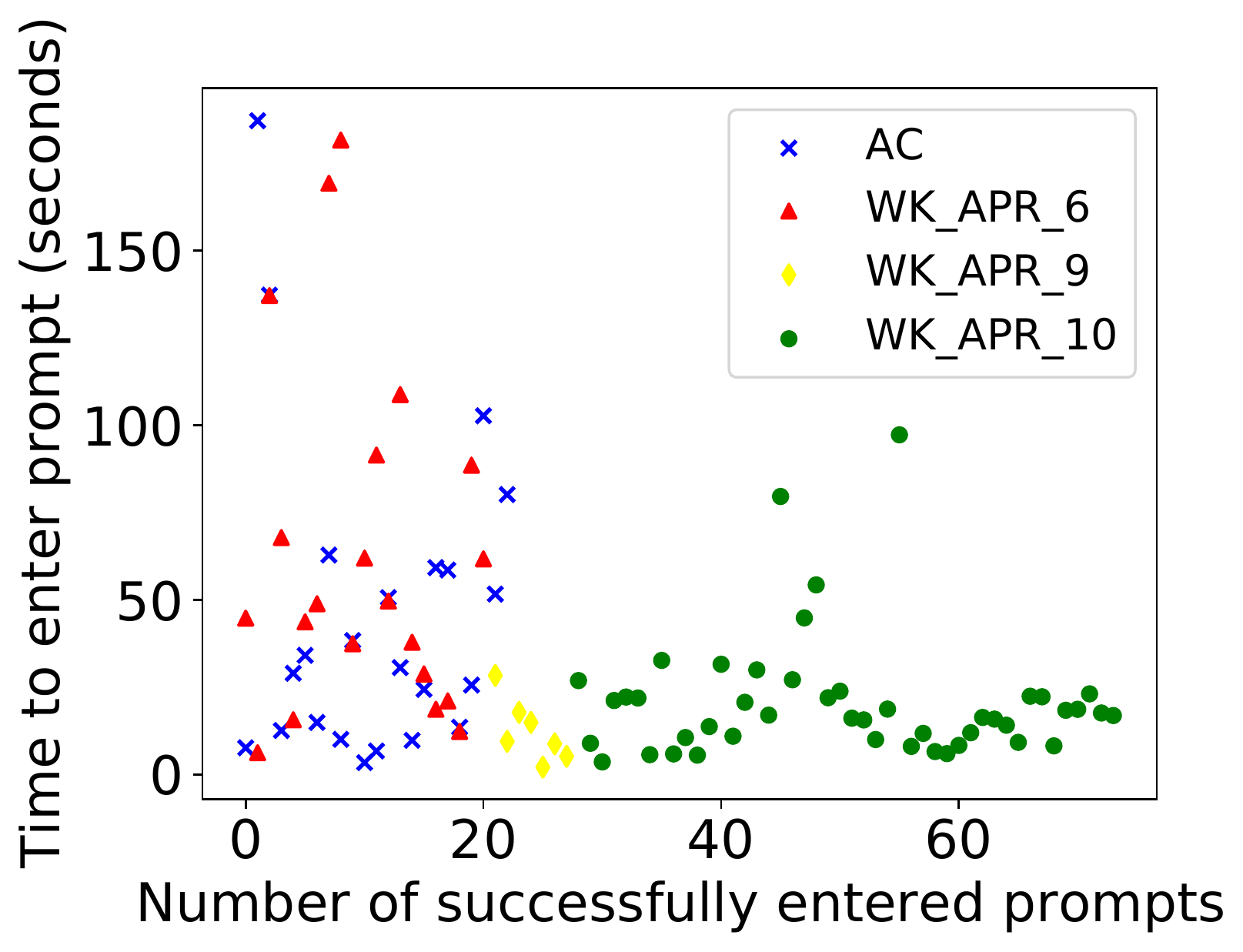}
\caption{Time taken by AC and WK to enter prompts using the MIO application}
\label{fig:timetoentervspromptno}
\end{figure}

Fig. \ref{fig:timetoentervspromptno} shows a scatter plot of the time taken to successfully enter a prompt against user experience. There is no clear trend in the users' first session with the application (blue crosses for WK, red triangles for AC). However, subsequent uses of the application (yellow diamonds and green circles for WK) indicate a significant decrease in the input time; analysis by linear regression gives us $\beta_1=-0.727$ $(p < 0.0002)$.

\section{Discussion and Future Work}

This paper aimed to introduce new means of interaction with computers for deafblind people by developing a vibrotactile interface. The involvement of the target users contrasts with the majority of previous work (Section \ref{section:related work}), which recruited sighted and hearing users, hence limiting the  applicability of their results in practice. An additional contribution is the adoption of user-centered design and the involvement of professionals and academics with prior experience in deafblind culture and assistive technology. Thus, the MIO design choices were informed by users' needs. 

Based on our findings, the advantages of the MIO application include: low cost, (relative) learnability, and simplicity. Moreover, the MIO interface requires only three buttons for text input, and can be integrated into any Android application. Hence, it is cheaper than Braille displays or specially-built haptic devices, although devices such as DB-Hand \cite{caporusso2017wearable} (with a manufacturing cost of around \$150) may become a viable option in the future.

The transmission bandwidth (200 ms per unit, or 30-35 characters per minute) is greater than that of Dual Tactile Morse-Braille \cite{arato2014teaching}, which transmitted 15 characters per minute. MIO is especially appropriate for short messages, consistent with \cite{plaisier2020learning}. Our results additionally confirm that the relative ease with which Morse was learned without visual reference by sighted participants in \cite{plaisier2020learning} is likely to extend to deafblind people with high literacy.

Furthermore, our findings show that the MIO interface was sufficiently easy to learn for a deafblind user who is proficient with other assistive technology, even if the user did not have experience with Morse code or haptic feedback. However, MIO's perceived usability is inconclusive. The SUS results indicated a wide gap in the perceived usability of the interface, which may be due to the difference in the two participants’ past experience. This study focused on measuring users' ability to recognize and input strings from a subset of the alphabet. 

Prior in-depth case studies for usability involving participants with disabilities typically had between three and 10 participants. This allowed for more generalizable findings \cite{lazar2017research}. However, similar to our research, the sole previous study that evaluated smartphone vibration-based haptics for the deafblind had only one deafblind participant \cite{arato2014teaching}. Future work might include an increased participant sample size and study duration (to allow more time for users to become familiar with the interface). 

In future work, longitudinal studies can be considered, comparing MIO to deafblind users’ usual text entry methods, after reaching similar familiarity with both methods. Moreover, MIO should be implemented on iOS devices to enable a greater numbers of deafblind participants, and the collection of interface usage data (e.g., usage time, rate of sending/receiving text). As per WK's recommendation, the application could be further extended to intercept notifications (e.g., Whatsapp and SMS messages), and vibrate the identity of the sender.

\section*{Acknowledgments}

We thank Heather Colson-Osborne (Anne Sullivan Foundation, Ireland); Mr. Scott Davert; Silvia Fajard Flores (Universidad de Colima); Dr. Marion Hersh (Glasgow University); Charmaine Tang (Singapore Association for the Deaf); Katie Trowsdale and Holly Whittome (proofreading); and the study's two participants, AC and WK, without whom there would be little to write about!

\bibliography{projectpaper}

\end{document}